\documentclass[letterpaper,amsmath,amssymb,prl,showpacs,superscriptaddress,twocolumn]{revtex4}

\usepackage{graphicx}
\usepackage{bm}

\usepackage{type1cm}
\usepackage{eso-pic}

\newcommand{\ket}[1]{\left| #1 \right>}

\begin{document}
\title{Optomechanical superpositions via nested interferometry}

\author{Brian Pepper}
\affiliation{Department of Physics, University of California, Santa Barbara, California 93106, USA}
\author{Roohollah Ghobadi}
\affiliation{Institute for Quantum Information Science and Department of Physics and Astronomy, University of Calgary, Calgary T2N 1N4, Alberta, Canada}
\affiliation{Department of Physics, Sharif University of Technology, Tehran, Iran}
\author{Evan Jeffrey}
\affiliation{Huygens Laboratory, Leiden University, P.O. Box 9504, 2300 RA Leiden, The Netherlands}
\author{Christoph Simon}
\affiliation{Institute for Quantum Information Science and Department of Physics and Astronomy, University of Calgary, Calgary T2N 1N4, Alberta, Canada}
\author{Dirk Bouwmeester}
\affiliation{Department of Physics, University of California, Santa Barbara, California 93106, USA}
\affiliation{Huygens Laboratory, Leiden University, P.O. Box 9504, 2300 RA Leiden, The Netherlands}

\date{\today}

\begin{abstract}
We present a scheme for achieving macroscopic quantum superpositions in optomechanical systems by using single photon postselection and detecting them with nested interferometers. This method relieves many of the challenges associated with previous optical schemes for measuring macroscopic superpositions, and only requires the devices to be in the weak coupling regime. It requires only small improvements on currently achievable device parameters, and allows observation of decoherence on a timescale unconstrained by the system's optical decay time. Prospects for observing novel decoherence mechanisms are discussed.
\end{abstract}

\maketitle


Optomechanical systems have been proposed as a method of achieving quantum superposition in mesoscopic systems \cite{Bose1999PRA, Marshall2003, Vitali2007}. However, such proposals impose several demanding experimental requirements, namely: a sideband-resolved cavity for ground state cooling \cite{Kleckner2008, Wilson-Rae2007, Marquardt2007, Teufel2011, Chan2011, Kleckner2011}, a coupling rate faster than the mechanical frequency in order to displace the mechanical state by more than its zero point fluctuation \cite{Marshall2003,Kleckner2008}, and strong optomechanical coupling to ensure photons remain in the cavity long enough to produce quantum effects \cite{Marshall2003,Thompson2008,Groblacher2009}. In practice, many of these requirements can be met individually, but they are extremely difficult to meet simultaneously. For instance, a recent result on diffraction-limited cavities \cite{Kleckner2010} has identified restrictions on achievable optical finesse in cavities with one micromirror end.

One approach to this challenge is to use coherent pumping to reach strong coupling in a device that would otherwise be weakly coupled \cite{Vitali2007,Thompson2008,Groblacher2009,Romero2010,Khalili2010,Akram2010}. This poses problems of its own, as it requires an elaborate readout scheme to distinguish a single photon from a large coherent background \cite{Akram2010} and is potentially vulnerable to laser phase noise \cite{Abdi2011,Ghobadi2011}. Another scheme uses levitated dielectric spheres \cite{Romero2011PRL} in the pulsed optomechanics regime \cite{Vanner2011}, but has stringent experimental requirements including extremely high vacuum ($10^{-16}$ torr) and may need to be performed in space \cite{Kaltenbaek2012}. Other quantum effects are also possible, such as squeezing the motion of the mechanical resonator via active feedback \cite{Clerk2008} or quadratic coupling \cite{Nunnenkamp2010}.

\begin{figure}[htbp]
\begin{center}
\includegraphics[width=4cm]{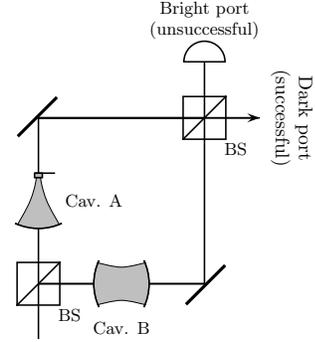}
\caption{The photon enters the first beam splitter of the inner interferometer, followed by an optomechanical cavity (A) and a conventional cavity (B). The photon weakly excites the optomechanical resonator. After the second beam splitter, dark port detection postselects for the case where the resonator has been excited by a phonon.}
\label{fig:postsel}
\end{center}
\end{figure}

In this paper, we propose using nested interferometers to create and detect macroscopic quantum superpositions. In the inner interferometer (see Fig.~\ref{fig:postsel}), we use postselection to amplify the effects of a single photon in the weak coupling regime. In the no-coupling limit photons always exit one port, and only when there is an optomechanical interaction can they be detected at the dark port. Postselecting dark port events results in distinguishable mechanical states (states with little overlap with ground state $\ket{0}_m$).

\begin{figure}[htbp]
\begin{center}
\includegraphics[width=8cm]{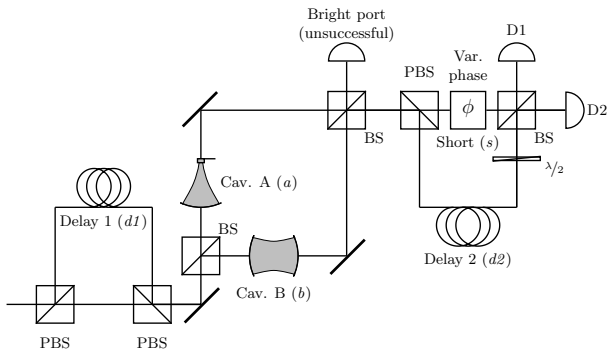}
\caption{Creating a macroscopic superposition via postselection and observing its decoherence. Note that though the two delay lines are depicted separately for clarity, in a realistic system the same delay line might be used twice, in each polarization mode.}
\label{fig:readout}
\end{center}
\end{figure}

Whereas ``dark port detection'' has already been proposed in the past, for example in the context of gravitational wave detection \cite{Weiss1972}, the main novelty of our scheme is that the inner interferometer is nested within a Franson \cite{Franson1991}, or time-bin \cite{Tittel1998}, interferometer (Fig.~\ref{fig:readout}). A single-photon state is split into a long and short path before entering the inner interferometer. In the dark port detection arm a second pair of long and short paths are present before the final detection of the photon on either detector D1 or D2. We will explain how the time-bin interferometry allows for the investigation of the coherent properties of the mechanical resonator in cavity A.

The Hamiltonian for optomechanical systems is given as follows \cite{Law1995}:
\begin{align}
\hat{\mathcal{H}}=\hbar\omega_o \hat{a}^{\dag}\hat{a} + \hbar\omega_m \hat{c}^{\dag}\hat{c} - \hbar g \hat{a}^{\dag}\hat{a}\left( \hat{c}+\hat{c}^{\dag}\right),\label{eqn:law}
\end{align}
where $\hbar$ is the reduced Planck's constant, $\omega_o$ is the optical angular frequency, $\hat{a}$ is the optical annihilation operator, $\omega_m$ is the mechanical angular frequency, $\hat{c}$ is the mechanical annihilation operator, and coupling strength $g=(\omega_o / L)\sqrt{\hbar / (2 m \omega_m)}$, with $L$ the cavity length and $m$ the effective mass of the mechanical mode.

A single photon in an optomechanical cavity interacts weakly with the mechanical mode, producing an periodic coherent displacement in the mechanical state \cite{Marshall2003, Bose1997PRA} of 
$\ket{\psi(t)}_m=\exp(i\phi(t)) \ket{\alpha(t)}_m$
with $\phi(t)=\kappa^2(\omega_m t - \sin \omega_m t)$, and $\alpha(t)=\kappa (1-e^{-i \omega_m t})$ and $\kappa=g/\omega_m$.  Since the interaction is weak, $\alpha(t) \ll 1$ at all times, making the displacement of the mechanical state hard to detect.


Now consider a Mach-Zehnder interferometer, where one arm contains an optomechanical cavity, and the other contains a stationary Fabry-P\'{e}rot cavity (with annihilation operator $\hat{b}$), as in \cite{Marshall2003}. This is shown in Fig.~\ref{fig:postsel}. The optomechanical device is cooled to the ground state using sideband-resolved cooling techniques \cite{Wilson-Rae2007,Marquardt2007}, and the cooling beam is switched off. A single photon is input to the interferometer, and after the first beam splitter the state of the system is $\ket{\psi_i} = \frac{1}{\sqrt{2}}\left(\ket{1}_{a}\ket{0}_{b} + \ket{0}_{a}\ket{1}_{b} \right)$. The photon weakly interacts with the optomechanical device, resulting in an overall state of:

\begin{align}
\ket{\psi}&=\frac{1}{\sqrt{2}}\Big[\ket{1}_{a}\ket{0}_{b}\ket{\psi(t)}_{m} + \ket{0}_{a}\ket{1}_{b}\ket{0}_{m} \Big]\nonumber\\
&\approx \frac{1}{\sqrt{2}}\left[e^{-|\alpha(t)|^{2} / 2}\left(\ket{1}_{a}\ket{0}_{b}\ket{0}_m + \right. \right. \nonumber\\
&\quad\quad\quad\quad\left. \alpha(t) \ket{1}_{a}\ket{0}_{b}\ket{1}_m \right) + \ket{0}_{a}\ket{1}_{b}\ket{0}_{m} \Big].
\end{align}

The second beam splitter postselects for an optical state $\ket{\psi_f}$ tuned such that the $\ket{0}_m$ components cancel each other out. Technically, this will vary depending on how long the photon remained in the cavity, but for $\alpha(t) \ll 1$ it will always be approximately $\ket{\psi_f} = \frac{1}{\sqrt{2}}\left(\ket{1}_{a}\ket{0}_{b} - \ket{0}_{a}\ket{1}_{b} \right)$. When photons exit the dark port of the interferometer, the state $\ket{\psi_f}$ is postselected, resulting in an unnormalized state of:

\begin{align}
\ket{\psi} \approx \left[\frac{e^{-|\alpha(t)|^{2}/2} - 1}{2}\ket{0}_m + \frac{\alpha(t)}{2} e^{-|\alpha(t)|^{2}/2} \ket{1}_m \right].
\end{align}
For $\alpha(t) \ll 1$, this is approximately $\ket{\psi} \approx (\alpha(t) / 2) \ket{1}_m$, or $\ket{1}_m$ with an $|\alpha(t)|^2 / 4$ chance of the postselection succeeding. We have thus probabilistically amplified the optomechanical effect of the photon.

This aspect of our scheme is related to the weak measurement formalism \cite{Aharonov1988,Aharonov1990}, with the optomechanical device essentially acting as a ``pointer'' which weakly measures photon number. However, it operates outside the weak measurement regime \cite{Geszti2010,Wu2011} due to its totally orthogonal postselection.

We propose to use this postselection to create macroscopic superpositions and measure their decoherence. Fig.~\ref{fig:readout} shows an extended optical setup, featuring an outer interferometer with two delay lines of equal length, one before the inner interferometer and one after it. The input photon is split by a polarizing beam splitter (PBS) into an early component and a late component which enters delay line 1. The early component immediately enters the inner interferometer and interacts with the device, and only the small component associated with mechanical state $\ket{1}_{m}$ passes through. After this component exits the dark port of the inner interferometer it is put into a second delay line via the polarizing beam splitter. At this point we have an entangled state, with a large component in delay line 1 associated with mechanical state $\ket{0}_{m}$, and a small component in delay line 2 associated with mechanical state $\ket{1}_{m}$. The late component then exits delay line 1 and enters the inner interferometer, where again only the component associated with $\ket{1}_{m}$ passes through. Finally, both components are interfered with each other at the end of the outer interferometer to check for visibility.

We sort the photons detected at the end of the outer interferometer into bins by arrival time. If the delay lines are of equal length $\tau_d$, then a photon detected at $t=\tau_d+t_{c}$ after the initial photon entered corresponds to a photon that remained in the cavities for time $t_{c}$. However, this conveys no information about whether it took the early or late path.  Thus, both components will have had the same value of $\alpha(t_{c})$, and both $\ket{1}_{m}$ components will have the same magnitude. Thus the early and late paths will be balanced and can interfere with perfect visibility.

Conditioned on the early component leaving the dark port of the inner interferometer, we will have an unnormalized state of:
\begin{align}
\ket{\psi}\approx\frac{1}{\sqrt{2}}\left(\ket{1}_{d1}\ket{0}_{d2}\ket{0}_m + \frac{\alpha(t_{c})}{2} \ket{0}_{d1}\ket{1}_{d2}\ket{1}_m \right),
\end{align}
with $d1$ and $d2$ labeling the first and second delay lines, respectively. This shows entanglement between the photon and the macroscopic mechanical state. Now, the components can be delayed for any length, optical losses allowing. After the late component has passed through the inner interferometer, we apply a variable phase $\phi$ to the early component, in order to observe fringes. Assuming no decoherence the state will be:
\begin{align}
\ket{\psi}\approx\frac{\alpha(t_{c})}{2\sqrt{2}}\left(e^{i\phi}\ket{1}_{s}\ket{0}_{d2}\ket{1}_m + \ket{0}_{s}\ket{1}_{d2}\ket{1}_m \right),\label{eqn:ps2}
\end{align}
with $s$ representing the short path of the late photon prior to the final beam splitter.

For increasing delay times, however, eventually the mechanical components will undergo decoherence of some kind. This could be traditional environmentally-induced decoherence due to imperfect isolation from the environment \cite{Zurek2003}, or it could be a proposed novel form of decoherence \cite{Ellis1984,Ellis1992,Ghirardi1990,Pearle1989,Penrose1996,Diosi1989PRA}. This would result in decay of the off-diagonal elements of $\ket{0}_m$ and $\ket{1}_m$.

After the final beamsplitter, there are two quantities that can be measured to characterize the superposition.

\begin{figure}[htbp]
\begin{center}
\includegraphics[width=8cm]{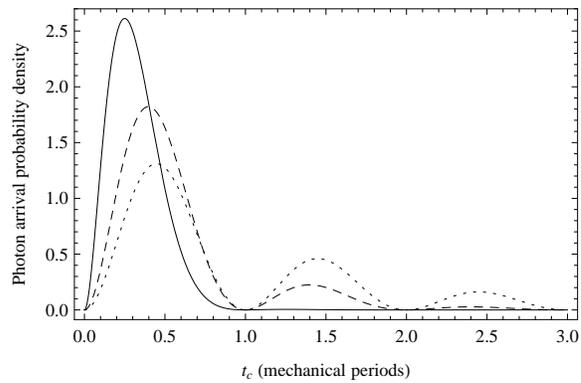}
\caption{Solid: probability density of a photon count vs. arrival time given a successful postselection for a sideband-resolved device with $\omega_m=\Gamma_c$. Dashed: $\omega_m=3\Gamma_c$. Dotted: $\omega_m=6\Gamma_c$.}
\label{fig:arrival}
\end{center}
\end{figure}

First, we can determine the arrival rate of photons versus time. Here we assume a single photon enters the cavity at a specific time, valid in the short-pulse limit \cite{Yang2011}. The probability density of a photon in a cavity being released after time $t_{c}$ is $\Gamma_{c}\exp(-\Gamma_{c} t_{c})$, where $\Gamma_{c}$ is the decay rate of the cavity. The probability of a successful postselection of a photon being released after $t_{c}$ is approximately $|\alpha(t_{c})|^2 / 4 = \kappa^2 \sin^2(\omega_m t_{c} / 2)$. Multiplying these results in a characteristic oscillation (Fig.~\ref{fig:arrival}) in arrival rate at the mechanical frequency of the optomechanical device. We can detect this oscillation by binning the photons by arrival time and comparing arrival rates. This indicates a successful postselection involving the device, ruling out counts on the dark port of an imperfectly aligned inner interferometer or entanglement with some other degree of freedom. Integrating, we get the overall probability of a single photon successfully creating a $\ket{1}_m$ state: 
\begin{align}
\kappa^2 \Gamma_{c}\int_0^\infty\sin^2\left(\frac{\omega_m t_{c}}{2}\right)e^{-\Gamma_{c} t_{c}} dt_{c} = \frac{1}{2}\frac{\kappa^2\omega_m^2}{\Gamma_{c}^2+\omega_m^2}.\label{eqn:integ}
\end{align}

Second, we can measure the interference visibility by varying the phase in the outer interferometer (``Var. phase'' in Fig.~\ref{fig:readout}). The visibility should not vary with arrival time in a given experiment; both components will have been put into the same mechanical state (Eqn.~\ref{eqn:ps2}). However, we can jointly vary the delay line lengths and plot visibility versus delay time. As delay time increases, visibility will eventually be lost due to some form of decoherence. Definitively determining the cause of any observed decoherence is difficult, but it will be possible to test its dependence on parameters like mass, frequency, environmental temperature, and mechanical Q, putting bounds on proposed macroscopic decoherence mechanisms.


\begin{table*}[htb]
\begin{center}
\begin{tabular}{l r@{}l r@{}l r@{}l r@{}l r@{}l c r@{}l r@{}l}
\multicolumn{1}{c}{Device} & \multicolumn{2}{c}{$m$ (ng)} & \multicolumn{2}{c}{$f_m$ (kHz)} & \multicolumn{2}{c}{$L$ (cm)} & \multicolumn{2}{c}{$F$} & \multicolumn{2}{c}{$Q_{m}$} & \multicolumn{1}{c}{$T_{\text{EID}}$ (K)} & \multicolumn{2}{c}{$\kappa$} & \multicolumn{2}{c}{$\omega_m / \Gamma_c$}\\
\hline
Trampoline resonator \#1 \cite{Kleckner2011} &  60 && 158 &     & 5 &     &    38,000 &&    43,000 && 0.3 & 0 & .000034 & 2 & .0 \\
Trampoline resonator \#2 \cite{Kleckner2011} & 110 &&   9 & .71 & 5 &     &    29,000 &&   940,000 && 0.4 & 0 & .0016   & 0 & .09 \\
Proposed device \#1                  &   1 && 300 &     & 0 & .5  &   300,000 &&    20,000 && 0.3 & 0 & .001    & 3 & .0 \\
Proposed device \#2                  &  100 &&   4 & .5 & 5 &     & 2,000,000 && 2,000,000 && 0.4 & 0 & .005    & 3 & .0 \\
\hline
\end{tabular}
\end{center}
\caption{Effective mass, mechanical frequency, cavity length, optical finesse, mechanical quality factor, environmentally induced decoherence temperature, and $\kappa=g/\omega_m$ of two recent devices with $\kappa$ high enough to attempt the proposed scheme. Trampoline resonator \#1 has insufficient $\kappa$ while \#2 has insufficient finesse to be sideband-resolved. Improved parameters for two devices with $\omega_m\simeq 3 \Gamma_c$ and $\kappa\simeq0.001$--$0.005$ are also presented. Proposed device \#2 could be used to observe novel decoherence mechanisms \cite{Ellis1984,Ellis1992,Penrose1996,Diosi1989PRA}.}
\label{tab:realdev}
\end{table*}

We now discuss this scheme's experimental requirements. First, the optomechanical device must be capable of cooling to the mechanical ground state. For the low-frequency devices considered here cooling by conventional means, as in \cite{OConnell2010}, is impractical. This means they must be in the sideband-resolved regime, $\omega_m\gtrsim\Gamma_c$, to allow optical ground state cooling \cite{Wilson-Rae2007,Marquardt2007}. Further, they must be a few times sideband-resolved, $\omega_m\gtrsim 3\Gamma_c$, in order to allow observation of the oscillations in arrival rate shown in Fig.~\ref{fig:arrival}. Many sideband-resolved devices \cite{Schliesser2007,Park2009,Schliesser2009,Teufel2011,Chan2011,Kleckner2011} have been demonstrated, and two have been successfully cooled to the ground state \cite{Teufel2011,Chan2011}.

The device must also have $\kappa$ high enough to make successful postselections common, though the precise value required will depend on the dark count rate of the detectors and the stability of the setup. As shown in Eqn.~\ref{eqn:integ}, a device with $\omega_m = 3\Gamma_c$ will have successful postselections with probability approximately $9\kappa^2/20$. The window in which the detectors will need to be open for photons is approximately $1/\Gamma_{c}$, leading to a requirement that the dark count rate be lower than $9\kappa^{2}\Gamma_c/20$.
The best silicon avalanche photodiodes (APDs) have a dark count rate of $\sim$2~Hz, requiring $\kappa\gtrsim 0.0009$ for a $300$~kHz device with $\omega_m=3\Gamma_c$, and $\kappa\gtrsim 0.007$ for a $4.5$~kHz device.

However, an emerging option is superconducting transition edge sensors (TESs) \cite{Lita2008}, which have negligible dark counts caused only by background thermal radiation \cite{Cabrera1998}. Dark counts this low would result in interferometer alignment being the limiting factor on $\kappa$. Though compared to APDs they have low maximum count rates ($\sim$100~kHz), poor time resolution ($\sim$0.1~$\mu$s) and require sub-Kelvin temperatures, none of these are problematic for the proposed experiment.

Table~\ref{tab:realdev} shows the parameters for two trampoline resonator devices \cite{Kleckner2011} representing the current state of the art, in terms of maximizing $\kappa$. It also shows two sets of proposed parameters representing devices with $\kappa\simeq 0.001$--$0.005$ and $\omega_m\simeq 3 \Gamma_c$, with only slight improvements over existing devices. The required finesse ranges from 300,000--2,000,000. For comparison, the highest reported finesse in an optical Fabry-P\'{e}rot cavity is $1.9\times 10^6$ \cite{Rempe1992}, and the highest reported between micromirrors is $1.5\times 10^5$ \cite{Muller2010}. This indicates that a sideband-resolved device with sufficient $\kappa$ for the proposed experiment is a realistic goal. For the proposed devices presented in Table~\ref{tab:realdev} it should be possible to collect a usable amount of data in times ranging from hours to days. These times depend on the specific device as well as the timescale of the decoherence being probed.

Further, the delay lines must be capable of storing the photons for multiple mechanical periods without significant losses. For delays up to $\sim$100~$\mu$s simple fiber optic delay lines are sufficient; at $1550$~nm fiber optic delay lines have acceptable losses (0.2 dB/km) for this purpose.  For shorter wavelengths fiber optic losses are too high but free space delay lines such as a Herriott cell may be used \cite{Herriott1965, Jeffrey2007, Vali73}, allowing $\sim$70~$\mu$s of delay. This could be increased to tens of milliseconds with ultrahigh reflectivity mirrors and very long cell lengths (lengths up to $1$~km have been demonstrated). In the future, much longer delay times may be possible using quantum optical memory \cite{Zhang2009,Radnaev2010}.

In addition, the base temperature from which optical cooling starts must be low enough that the ground state can withstand environmentally induced decoherence for multiple mechanical periods. This requirement is given as $T\ll T_{\text{EID}}\equiv\hbar\omega_{m} Q_{m}/k_{B}$ \cite{Zurek2003,Marshall2003,Kleckner2008}. This means that mechanical quality factor $Q_m$ must be high enough that it is possible to cool below $T_{\text{EID}}$ prior to optical cooling. The values of $T_{\text{EID}}$ for the devices in Table~\ref{tab:realdev} are easily met by a standard dilution refrigerator.

It is important to note that the proposed scheme is potentially useful for other types of weakly coupled optomechanical devices, even in very different frequency regimes. For instance, optomechanical crystals with $\omega_m=7.4\Gamma_c=2\pi\times 3.68$~GHz and $\kappa=0.00025$ have been demonstrated \cite{Chan2011}. Though the lower value of $\kappa$ lowers the chance of a successful postselection and places stricter requirements on the alignment of the inner interferometer, the higher frequency might allow experimental runs to be performed in similar amounts of time.

We can explore the decoherence timescales predicted by various novel decoherence schemes, using the proposed devices from Table~\ref{tab:realdev}. For quantum gravitational collapse, following \cite{Ellis1992,Romero2011}, we find decoherence timescales of order $10$~s for proposed device \#1 and $1$~ms for proposed device \#2, possibly testable with the proposed scheme. For continuous spontaneous localization, following \cite{Collett2003,Romero2011}, we find decoherence timescales of order $10^{7}$~s and $10^{5}$~s respectively, out of reach for our scheme. For a test of gravitationally induced decoherence \cite{Penrose1996, Diosi1989PRA}, the matter is more complicated, as there is considerable theoretical disagreement about what mass distribution to use for the nuclei of the system \cite{Diosi2007,Kleckner2008,Maimone2011,Romero2011}. Regarding the nuclei as having sizes equal to their zero point motion in the lattice results in decoherence times on the order of $10^{6}$~s and $10^{4}$~s respectively, out of reach for our scheme. More optimistically, regarding their size as the size of the atomic nuclei as in \cite{Kleckner2008} would result in decoherence times on the order of $10$~ms and $100$~$\mu$s respectively, testable with the current scheme.

For comparison, for the proposed devices at a base temperature of $1$~mK, we would expect environmentally induced decoherence from coupling to the bath \cite{Zurek2003} to have decoherence times of $\sim$150~$\mu$s and $\sim$15~ms respectively.



In conclusion, we have proposed a method of postselected nested interferometry for the creation and investigation of macroscopic quantum superpositions. This scheme has two notable advantages over previous optical schemes \cite{Marshall2003}: it only requires weakly coupled optomechanical systems, and the mechanical decoherence times that can be investigated are not limited by the optical storage time within the optomechanical system but only by the optical storage time in external delays. As a result, it is realizable with only slight improvements over existing devices.


The authors gratefully acknowledge support by the National Science Foundation grant PHY-0804177, Marie-Curie EXT-CT-2006-042580, European Commission Project MINOS, NWO VICI grant 680-47-604, an AITF New Faculty Award, and an NSERC Discovery Grant.


\begin{thebibliography}{56}
\expandafter\ifx\csname natexlab\endcsname\relax\def\natexlab#1{#1}\fi
\expandafter\ifx\csname bibnamefont\endcsname\relax
  \def\bibnamefont#1{#1}\fi
\expandafter\ifx\csname bibfnamefont\endcsname\relax
  \def\bibfnamefont#1{#1}\fi
\expandafter\ifx\csname citenamefont\endcsname\relax
  \def\citenamefont#1{#1}\fi
\expandafter\ifx\csname url\endcsname\relax
  \def\url#1{\texttt{#1}}\fi
\expandafter\ifx\csname urlprefix\endcsname\relax\def\urlprefix{URL }\fi
\providecommand{\bibinfo}[2]{#2}
\providecommand{\eprint}[2][]{\url{#2}}

\bibitem[{\citenamefont{{Bose} et~al.}(1999)\citenamefont{{Bose}, {Jacobs}, and
  {Knight}}}]{Bose1999PRA}
\bibinfo{author}{\bibfnamefont{S.}~\bibnamefont{{Bose}}},
  \bibinfo{author}{\bibfnamefont{K.}~\bibnamefont{{Jacobs}}}, \bibnamefont{and}
  \bibinfo{author}{\bibfnamefont{P.~L.} \bibnamefont{{Knight}}},
  \bibinfo{journal}{Phys. Rev. A} \textbf{\bibinfo{volume}{59}},
  \bibinfo{pages}{3204} (\bibinfo{year}{1999}).

\bibitem[{\citenamefont{Marshall et~al.}(2003)\citenamefont{Marshall, Simon,
  Penrose, and Bouwmeester}}]{Marshall2003}
\bibinfo{author}{\bibfnamefont{W.}~\bibnamefont{Marshall}},
  \bibinfo{author}{\bibfnamefont{C.}~\bibnamefont{Simon}},
  \bibinfo{author}{\bibfnamefont{R.}~\bibnamefont{Penrose}}, \bibnamefont{and}
  \bibinfo{author}{\bibfnamefont{D.}~\bibnamefont{Bouwmeester}},
  \bibinfo{journal}{Phys. Rev. Lett.} \textbf{\bibinfo{volume}{91}},
  \bibinfo{pages}{130401} (\bibinfo{year}{2003}).

\bibitem[{\citenamefont{Vitali et~al.}(2007)\citenamefont{Vitali, Gigan,
  Ferreira, B\"ohm, Tombesi, Guerreiro, Vedral, Zeilinger, and
  Aspelmeyer}}]{Vitali2007}
\bibinfo{author}{\bibfnamefont{D.}~\bibnamefont{Vitali}},
  \bibinfo{author}{\bibfnamefont{S.}~\bibnamefont{Gigan}},
  \bibinfo{author}{\bibfnamefont{A.}~\bibnamefont{Ferreira}},
  \bibinfo{author}{\bibfnamefont{H.~R.} \bibnamefont{B\"ohm}},
  \bibinfo{author}{\bibfnamefont{P.}~\bibnamefont{Tombesi}},
  \bibinfo{author}{\bibfnamefont{A.}~\bibnamefont{Guerreiro}},
  \bibinfo{author}{\bibfnamefont{V.}~\bibnamefont{Vedral}},
  \bibinfo{author}{\bibfnamefont{A.}~\bibnamefont{Zeilinger}},
  \bibnamefont{and}
  \bibinfo{author}{\bibfnamefont{M.}~\bibnamefont{Aspelmeyer}},
  \bibinfo{journal}{Phys. Rev. Lett.} \textbf{\bibinfo{volume}{98}},
  \bibinfo{pages}{030405} (\bibinfo{year}{2007}).

\bibitem[{\citenamefont{{Kleckner} et~al.}(2008)\citenamefont{{Kleckner},
  {Pikovski}, {Jeffrey}, {Ament}, {Eliel}, {van den Brink}, and
  {Bouwmeester}}}]{Kleckner2008}
\bibinfo{author}{\bibfnamefont{D.}~\bibnamefont{{Kleckner}}},
  \bibinfo{author}{\bibfnamefont{I.}~\bibnamefont{{Pikovski}}},
  \bibinfo{author}{\bibfnamefont{E.}~\bibnamefont{{Jeffrey}}},
  \bibinfo{author}{\bibfnamefont{L.}~\bibnamefont{{Ament}}},
  \bibinfo{author}{\bibfnamefont{E.}~\bibnamefont{{Eliel}}},
  \bibinfo{author}{\bibfnamefont{J.}~\bibnamefont{{van den Brink}}},
  \bibnamefont{and}
  \bibinfo{author}{\bibfnamefont{D.}~\bibnamefont{{Bouwmeester}}},
  \bibinfo{journal}{New J. Phys.} \textbf{\bibinfo{volume}{10}},
  \bibinfo{pages}{095020} (\bibinfo{year}{2008}).

\bibitem[{\citenamefont{Wilson-Rae et~al.}(2007)\citenamefont{Wilson-Rae,
  Nooshi, Zwerger, and Kippenberg}}]{Wilson-Rae2007}
\bibinfo{author}{\bibfnamefont{I.}~\bibnamefont{Wilson-Rae}},
  \bibinfo{author}{\bibfnamefont{N.}~\bibnamefont{Nooshi}},
  \bibinfo{author}{\bibfnamefont{W.}~\bibnamefont{Zwerger}}, \bibnamefont{and}
  \bibinfo{author}{\bibfnamefont{T.~J.} \bibnamefont{Kippenberg}},
  \bibinfo{journal}{Phys. Rev. Lett.} \textbf{\bibinfo{volume}{99}},
  \bibinfo{pages}{093901} (\bibinfo{year}{2007}).

\bibitem[{\citenamefont{Marquardt et~al.}(2007)\citenamefont{Marquardt, Chen,
  Clerk, and Girvin}}]{Marquardt2007}
\bibinfo{author}{\bibfnamefont{F.}~\bibnamefont{Marquardt}},
  \bibinfo{author}{\bibfnamefont{J.~P.} \bibnamefont{Chen}},
  \bibinfo{author}{\bibfnamefont{A.~A.} \bibnamefont{Clerk}}, \bibnamefont{and}
  \bibinfo{author}{\bibfnamefont{S.~M.} \bibnamefont{Girvin}},
  \bibinfo{journal}{Phys. Rev. Lett.} \textbf{\bibinfo{volume}{99}},
  \bibinfo{pages}{093902} (\bibinfo{year}{2007}).

\bibitem[{\citenamefont{{Teufel} et~al.}(2011)\citenamefont{{Teufel}, {Donner},
  {Li}, {Harlow}, {Allman}, {Cicak}, {Sirois}, {Whittaker}, {Lehnert}, and
  {Simmonds}}}]{Teufel2011}
\bibinfo{author}{\bibfnamefont{J.~D.} \bibnamefont{{Teufel}}},
  \bibinfo{author}{\bibfnamefont{T.}~\bibnamefont{{Donner}}},
  \bibinfo{author}{\bibfnamefont{D.}~\bibnamefont{{Li}}},
  \bibinfo{author}{\bibfnamefont{J.~W.} \bibnamefont{{Harlow}}},
  \bibinfo{author}{\bibfnamefont{M.~S.} \bibnamefont{{Allman}}},
  \bibinfo{author}{\bibfnamefont{K.}~\bibnamefont{{Cicak}}},
  \bibinfo{author}{\bibfnamefont{A.~J.} \bibnamefont{{Sirois}}},
  \bibinfo{author}{\bibfnamefont{J.~D.} \bibnamefont{{Whittaker}}},
  \bibinfo{author}{\bibfnamefont{K.~W.} \bibnamefont{{Lehnert}}},
  \bibnamefont{and} \bibinfo{author}{\bibfnamefont{R.~W.}
  \bibnamefont{{Simmonds}}}, \bibinfo{journal}{Nature}
  \textbf{\bibinfo{volume}{475}}, \bibinfo{pages}{359} (\bibinfo{year}{2011}).

\bibitem[{\citenamefont{Chan et~al.}(2011)\citenamefont{Chan, Alegre,
  Safavi-Naeini, Hill, Krause, Groblacher, Aspelmeyer, and Painter}}]{Chan2011}
\bibinfo{author}{\bibfnamefont{J.}~\bibnamefont{Chan}},
  \bibinfo{author}{\bibfnamefont{T.~P.~M.} \bibnamefont{Alegre}},
  \bibinfo{author}{\bibfnamefont{A.~H.} \bibnamefont{Safavi-Naeini}},
  \bibinfo{author}{\bibfnamefont{J.~T.} \bibnamefont{Hill}},
  \bibinfo{author}{\bibfnamefont{A.}~\bibnamefont{Krause}},
  \bibinfo{author}{\bibfnamefont{S.}~\bibnamefont{Groblacher}},
  \bibinfo{author}{\bibfnamefont{M.}~\bibnamefont{Aspelmeyer}},
  \bibnamefont{and} \bibinfo{author}{\bibfnamefont{O.}~\bibnamefont{Painter}},
  \bibinfo{journal}{Nature} \textbf{\bibinfo{volume}{478}}, \bibinfo{pages}{89}
  (\bibinfo{year}{2011}).

\bibitem[{\citenamefont{Kleckner et~al.}(2011)\citenamefont{Kleckner, Pepper,
  Jeffrey, Sonin, Thon, and Bouwmeester}}]{Kleckner2011}
\bibinfo{author}{\bibfnamefont{D.}~\bibnamefont{Kleckner}},
  \bibinfo{author}{\bibfnamefont{B.}~\bibnamefont{Pepper}},
  \bibinfo{author}{\bibfnamefont{E.}~\bibnamefont{Jeffrey}},
  \bibinfo{author}{\bibfnamefont{P.}~\bibnamefont{Sonin}},
  \bibinfo{author}{\bibfnamefont{S.~M.} \bibnamefont{Thon}}, \bibnamefont{and}
  \bibinfo{author}{\bibfnamefont{D.}~\bibnamefont{Bouwmeester}},
  \bibinfo{journal}{Opt. Express} \textbf{\bibinfo{volume}{19}},
  \bibinfo{pages}{19708} (\bibinfo{year}{2011}).

\bibitem[{\citenamefont{{Thompson} et~al.}(2008)\citenamefont{{Thompson},
  {Zwickl}, {Jayich}, {Marquardt}, {Girvin}, and {Harris}}}]{Thompson2008}
\bibinfo{author}{\bibfnamefont{J.~D.} \bibnamefont{{Thompson}}},
  \bibinfo{author}{\bibfnamefont{B.~M.} \bibnamefont{{Zwickl}}},
  \bibinfo{author}{\bibfnamefont{A.~M.} \bibnamefont{{Jayich}}},
  \bibinfo{author}{\bibfnamefont{F.}~\bibnamefont{{Marquardt}}},
  \bibinfo{author}{\bibfnamefont{S.~M.} \bibnamefont{{Girvin}}},
  \bibnamefont{and} \bibinfo{author}{\bibfnamefont{J.~G.~E.}
  \bibnamefont{{Harris}}}, \bibinfo{journal}{Nature}
  \textbf{\bibinfo{volume}{452}}, \bibinfo{pages}{72} (\bibinfo{year}{2008}).

\bibitem[{\citenamefont{Gr{\"o}blacher
  et~al.}(2009)\citenamefont{Gr{\"o}blacher, Hammerer, Vanner, and
  Aspelmeyer}}]{Groblacher2009}
\bibinfo{author}{\bibfnamefont{S.}~\bibnamefont{Gr{\"o}blacher}},
  \bibinfo{author}{\bibfnamefont{K.}~\bibnamefont{Hammerer}},
  \bibinfo{author}{\bibfnamefont{M.~R.} \bibnamefont{Vanner}},
  \bibnamefont{and}
  \bibinfo{author}{\bibfnamefont{M.}~\bibnamefont{Aspelmeyer}},
  \bibinfo{journal}{Nature} \textbf{\bibinfo{volume}{460}},
  \bibinfo{pages}{724} (\bibinfo{year}{2009}).

\bibitem[{\citenamefont{Kleckner et~al.}(2010)\citenamefont{Kleckner, Irvine,
  Oemrawsingh, and Bouwmeester}}]{Kleckner2010}
\bibinfo{author}{\bibfnamefont{D.}~\bibnamefont{Kleckner}},
  \bibinfo{author}{\bibfnamefont{W.~T.~M.} \bibnamefont{Irvine}},
  \bibinfo{author}{\bibfnamefont{S.~S.~R.} \bibnamefont{Oemrawsingh}},
  \bibnamefont{and}
  \bibinfo{author}{\bibfnamefont{D.}~\bibnamefont{Bouwmeester}},
  \bibinfo{journal}{Phys. Rev. A} \textbf{\bibinfo{volume}{81}},
  \bibinfo{pages}{043814} (\bibinfo{year}{2010}).

\bibitem[{\citenamefont{Romero-Isart et~al.}(2010)\citenamefont{Romero-Isart,
  Juan, Quidant, and Cirac}}]{Romero2010}
\bibinfo{author}{\bibfnamefont{O.}~\bibnamefont{Romero-Isart}},
  \bibinfo{author}{\bibfnamefont{M.~L.} \bibnamefont{Juan}},
  \bibinfo{author}{\bibfnamefont{R.}~\bibnamefont{Quidant}}, \bibnamefont{and}
  \bibinfo{author}{\bibfnamefont{J.~I.} \bibnamefont{Cirac}},
  \bibinfo{journal}{New J. Phys.} \textbf{\bibinfo{volume}{12}},
  \bibinfo{pages}{033015} (\bibinfo{year}{2010}).

\bibitem[{\citenamefont{Khalili et~al.}(2010)\citenamefont{Khalili, Danilishin,
  Miao, M\"uller-Ebhardt, Yang, and Chen}}]{Khalili2010}
\bibinfo{author}{\bibfnamefont{F.}~\bibnamefont{Khalili}},
  \bibinfo{author}{\bibfnamefont{S.}~\bibnamefont{Danilishin}},
  \bibinfo{author}{\bibfnamefont{H.}~\bibnamefont{Miao}},
  \bibinfo{author}{\bibfnamefont{H.}~\bibnamefont{M\"uller-Ebhardt}},
  \bibinfo{author}{\bibfnamefont{H.}~\bibnamefont{Yang}}, \bibnamefont{and}
  \bibinfo{author}{\bibfnamefont{Y.}~\bibnamefont{Chen}},
  \bibinfo{journal}{Phys. Rev. Lett.} \textbf{\bibinfo{volume}{105}},
  \bibinfo{pages}{070403} (\bibinfo{year}{2010}).

\bibitem[{\citenamefont{Akram et~al.}(2010)\citenamefont{Akram, Kiesel,
  Aspelmeyer, and Milburn}}]{Akram2010}
\bibinfo{author}{\bibfnamefont{U.}~\bibnamefont{Akram}},
  \bibinfo{author}{\bibfnamefont{N.}~\bibnamefont{Kiesel}},
  \bibinfo{author}{\bibfnamefont{M.}~\bibnamefont{Aspelmeyer}},
  \bibnamefont{and} \bibinfo{author}{\bibfnamefont{G.~J.}
  \bibnamefont{Milburn}}, \bibinfo{journal}{New J. Phys.}
  \textbf{\bibinfo{volume}{12}}, \bibinfo{pages}{083030}
  (\bibinfo{year}{2010}).

\bibitem[{\citenamefont{Abdi et~al.}(2011)\citenamefont{Abdi, Barzanjeh,
  Tombesi, and Vitali}}]{Abdi2011}
\bibinfo{author}{\bibfnamefont{M.}~\bibnamefont{Abdi}},
  \bibinfo{author}{\bibfnamefont{S.}~\bibnamefont{Barzanjeh}},
  \bibinfo{author}{\bibfnamefont{P.}~\bibnamefont{Tombesi}}, \bibnamefont{and}
  \bibinfo{author}{\bibfnamefont{D.}~\bibnamefont{Vitali}},
  \bibinfo{journal}{Phys. Rev. A} \textbf{\bibinfo{volume}{84}},
  \bibinfo{pages}{032325} (\bibinfo{year}{2011}).

\bibitem[{\citenamefont{Ghobadi et~al.}(2011)\citenamefont{Ghobadi, Bahrampour,
  and Simon}}]{Ghobadi2011}
\bibinfo{author}{\bibfnamefont{R.}~\bibnamefont{Ghobadi}},
  \bibinfo{author}{\bibfnamefont{A.~R.} \bibnamefont{Bahrampour}},
  \bibnamefont{and} \bibinfo{author}{\bibfnamefont{C.}~\bibnamefont{Simon}},
  \bibinfo{journal}{Phys. Rev. A} \textbf{\bibinfo{volume}{84}},
  \bibinfo{pages}{063827} (\bibinfo{year}{2011}).

\bibitem[{\citenamefont{Romero-Isart et~al.}(2011)\citenamefont{Romero-Isart,
  Pflanzer, Blaser, Kaltenbaek, Kiesel, Aspelmeyer, and Cirac}}]{Romero2011PRL}
\bibinfo{author}{\bibfnamefont{O.}~\bibnamefont{Romero-Isart}},
  \bibinfo{author}{\bibfnamefont{A.~C.} \bibnamefont{Pflanzer}},
  \bibinfo{author}{\bibfnamefont{F.}~\bibnamefont{Blaser}},
  \bibinfo{author}{\bibfnamefont{R.}~\bibnamefont{Kaltenbaek}},
  \bibinfo{author}{\bibfnamefont{N.}~\bibnamefont{Kiesel}},
  \bibinfo{author}{\bibfnamefont{M.}~\bibnamefont{Aspelmeyer}},
  \bibnamefont{and} \bibinfo{author}{\bibfnamefont{J.~I.} \bibnamefont{Cirac}},
  \bibinfo{journal}{Phys. Rev. Lett.} \textbf{\bibinfo{volume}{107}},
  \bibinfo{pages}{020405} (\bibinfo{year}{2011}).

\bibitem[{\citenamefont{Vanner et~al.}(2011)\citenamefont{Vanner, Pikovski,
  Cole, Kim, Brukner, Hammerer, Milburn, and Aspelmeyer}}]{Vanner2011}
\bibinfo{author}{\bibfnamefont{M.~R.} \bibnamefont{Vanner}},
  \bibinfo{author}{\bibfnamefont{I.}~\bibnamefont{Pikovski}},
  \bibinfo{author}{\bibfnamefont{G.~D.} \bibnamefont{Cole}},
  \bibinfo{author}{\bibfnamefont{M.~S.} \bibnamefont{Kim}},
  \bibinfo{author}{\bibfnamefont{{\v{C}}.}~\bibnamefont{Brukner}},
  \bibinfo{author}{\bibfnamefont{K.}~\bibnamefont{Hammerer}},
  \bibinfo{author}{\bibfnamefont{G.~J.} \bibnamefont{Milburn}},
  \bibnamefont{and}
  \bibinfo{author}{\bibfnamefont{M.}~\bibnamefont{Aspelmeyer}},
  \bibinfo{journal}{PNAS} \textbf{\bibinfo{volume}{108}},
  \bibinfo{pages}{16182} (\bibinfo{year}{2011}).

\bibitem[{\citenamefont{Kaltenbaek et~al.}(2012)\citenamefont{Kaltenbaek,
  Hechenblaikner, Kiesel, Romero-Isart, Schwab, Johann, and
  Aspelmeyer}}]{Kaltenbaek2012}
\bibinfo{author}{\bibfnamefont{R.}~\bibnamefont{Kaltenbaek}},
  \bibinfo{author}{\bibfnamefont{G.}~\bibnamefont{Hechenblaikner}},
  \bibinfo{author}{\bibfnamefont{N.}~\bibnamefont{Kiesel}},
  \bibinfo{author}{\bibfnamefont{O.}~\bibnamefont{Romero-Isart}},
  \bibinfo{author}{\bibfnamefont{K.}~\bibnamefont{Schwab}},
  \bibinfo{author}{\bibfnamefont{U.}~\bibnamefont{Johann}}, \bibnamefont{and}
  \bibinfo{author}{\bibfnamefont{M.}~\bibnamefont{Aspelmeyer}},
  \bibinfo{journal}{Exp. Astron.} pp. \bibinfo{pages}{1--42}
  (\bibinfo{year}{2012}).

\bibitem[{\citenamefont{Clerk et~al.}(2008)\citenamefont{Clerk, Marquardt, and
  Jacobs}}]{Clerk2008}
\bibinfo{author}{\bibfnamefont{A.~A.} \bibnamefont{Clerk}},
  \bibinfo{author}{\bibfnamefont{F.}~\bibnamefont{Marquardt}},
  \bibnamefont{and} \bibinfo{author}{\bibfnamefont{K.}~\bibnamefont{Jacobs}},
  \bibinfo{journal}{New J. Phys.} \textbf{\bibinfo{volume}{10}},
  \bibinfo{pages}{095010} (\bibinfo{year}{2008}).

\bibitem[{\citenamefont{Nunnenkamp et~al.}(2010)\citenamefont{Nunnenkamp,
  B\o{}rkje, Harris, and Girvin}}]{Nunnenkamp2010}
\bibinfo{author}{\bibfnamefont{A.}~\bibnamefont{Nunnenkamp}},
  \bibinfo{author}{\bibfnamefont{K.}~\bibnamefont{B\o{}rkje}},
  \bibinfo{author}{\bibfnamefont{J.~G.~E.} \bibnamefont{Harris}},
  \bibnamefont{and} \bibinfo{author}{\bibfnamefont{S.~M.}
  \bibnamefont{Girvin}}, \bibinfo{journal}{Phys. Rev. A}
  \textbf{\bibinfo{volume}{82}}, \bibinfo{pages}{021806}
  (\bibinfo{year}{2010}).

\bibitem[{\citenamefont{Weiss}(1972)}]{Weiss1972}
\bibinfo{author}{\bibfnamefont{R.}~\bibnamefont{Weiss}}, \bibinfo{journal}{MIT
  Res. Lab. Electron. Q. Prog. Rep.} \textbf{\bibinfo{volume}{105}},
  \bibinfo{pages}{54} (\bibinfo{year}{1972}).

\bibitem[{\citenamefont{Franson}(1991)}]{Franson1991}
\bibinfo{author}{\bibfnamefont{J.~D.} \bibnamefont{Franson}},
  \bibinfo{journal}{Phys. Rev. A} \textbf{\bibinfo{volume}{44}},
  \bibinfo{pages}{4552} (\bibinfo{year}{1991}).

\bibitem[{\citenamefont{Tittel et~al.}(1998)\citenamefont{Tittel, Brendel,
  Zbinden, and Gisin}}]{Tittel1998}
\bibinfo{author}{\bibfnamefont{W.}~\bibnamefont{Tittel}},
  \bibinfo{author}{\bibfnamefont{J.}~\bibnamefont{Brendel}},
  \bibinfo{author}{\bibfnamefont{H.}~\bibnamefont{Zbinden}}, \bibnamefont{and}
  \bibinfo{author}{\bibfnamefont{N.}~\bibnamefont{Gisin}},
  \bibinfo{journal}{Phys. Rev. Lett.} \textbf{\bibinfo{volume}{81}},
  \bibinfo{pages}{3563} (\bibinfo{year}{1998}).

\bibitem[{\citenamefont{Law}(1995)}]{Law1995}
\bibinfo{author}{\bibfnamefont{C.~K.} \bibnamefont{Law}},
  \bibinfo{journal}{Phys. Rev. A} \textbf{\bibinfo{volume}{51}},
  \bibinfo{pages}{2537} (\bibinfo{year}{1995}).

\bibitem[{\citenamefont{{Bose} et~al.}(1997)\citenamefont{{Bose}, {Jacobs}, and
  {Knight}}}]{Bose1997PRA}
\bibinfo{author}{\bibfnamefont{S.}~\bibnamefont{{Bose}}},
  \bibinfo{author}{\bibfnamefont{K.}~\bibnamefont{{Jacobs}}}, \bibnamefont{and}
  \bibinfo{author}{\bibfnamefont{P.~L.} \bibnamefont{{Knight}}},
  \bibinfo{journal}{Phys. Rev. A} \textbf{\bibinfo{volume}{56}},
  \bibinfo{pages}{4175} (\bibinfo{year}{1997}).

\bibitem[{\citenamefont{Aharonov et~al.}(1988)\citenamefont{Aharonov, Albert,
  and Vaidman}}]{Aharonov1988}
\bibinfo{author}{\bibfnamefont{Y.}~\bibnamefont{Aharonov}},
  \bibinfo{author}{\bibfnamefont{D.~Z.} \bibnamefont{Albert}},
  \bibnamefont{and} \bibinfo{author}{\bibfnamefont{L.}~\bibnamefont{Vaidman}},
  \bibinfo{journal}{Phys. Rev. Lett.} \textbf{\bibinfo{volume}{60}},
  \bibinfo{pages}{1351} (\bibinfo{year}{1988}).

\bibitem[{\citenamefont{Aharonov and Vaidman}(1990)}]{Aharonov1990}
\bibinfo{author}{\bibfnamefont{Y.}~\bibnamefont{Aharonov}} \bibnamefont{and}
  \bibinfo{author}{\bibfnamefont{L.}~\bibnamefont{Vaidman}},
  \bibinfo{journal}{Phys. Rev. A} \textbf{\bibinfo{volume}{41}},
  \bibinfo{pages}{11} (\bibinfo{year}{1990}).

\bibitem[{\citenamefont{Geszti}(2010)}]{Geszti2010}
\bibinfo{author}{\bibfnamefont{T.}~\bibnamefont{Geszti}},
  \bibinfo{journal}{Phys. Rev. A} \textbf{\bibinfo{volume}{81}},
  \bibinfo{pages}{044102} (\bibinfo{year}{2010}).

\bibitem[{\citenamefont{Wu and Li}(2011)}]{Wu2011}
\bibinfo{author}{\bibfnamefont{S.}~\bibnamefont{Wu}} \bibnamefont{and}
  \bibinfo{author}{\bibfnamefont{Y.}~\bibnamefont{Li}}, \bibinfo{journal}{Phys.
  Rev. A} \textbf{\bibinfo{volume}{83}}, \bibinfo{pages}{052106}
  (\bibinfo{year}{2011}).

\bibitem[{\citenamefont{Zurek}(2003)}]{Zurek2003}
\bibinfo{author}{\bibfnamefont{W.~H.} \bibnamefont{Zurek}},
  \bibinfo{journal}{Rev. Mod. Phys.} \textbf{\bibinfo{volume}{75}},
  \bibinfo{pages}{715} (\bibinfo{year}{2003}).

\bibitem[{\citenamefont{Ellis et~al.}(1984)\citenamefont{Ellis, Hagelin,
  Nanopoulos, and Srednicki}}]{Ellis1984}
\bibinfo{author}{\bibfnamefont{J.}~\bibnamefont{Ellis}},
  \bibinfo{author}{\bibfnamefont{J.~S.} \bibnamefont{Hagelin}},
  \bibinfo{author}{\bibfnamefont{D.}~\bibnamefont{Nanopoulos}},
  \bibnamefont{and}
  \bibinfo{author}{\bibfnamefont{M.}~\bibnamefont{Srednicki}},
  \bibinfo{journal}{Nucl. Phys. B} \textbf{\bibinfo{volume}{241}},
  \bibinfo{pages}{381 } (\bibinfo{year}{1984}).

\bibitem[{\citenamefont{Ellis et~al.}(1992)\citenamefont{Ellis, Mavromatos, and
  Nanopoulos}}]{Ellis1992}
\bibinfo{author}{\bibfnamefont{J.}~\bibnamefont{Ellis}},
  \bibinfo{author}{\bibfnamefont{N.}~\bibnamefont{Mavromatos}},
  \bibnamefont{and}
  \bibinfo{author}{\bibfnamefont{D.}~\bibnamefont{Nanopoulos}},
  \bibinfo{journal}{Phys. Lett. B} \textbf{\bibinfo{volume}{293}},
  \bibinfo{pages}{37 } (\bibinfo{year}{1992}).

\bibitem[{\citenamefont{Ghirardi et~al.}(1990)\citenamefont{Ghirardi, Pearle,
  and Rimini}}]{Ghirardi1990}
\bibinfo{author}{\bibfnamefont{G.~C.} \bibnamefont{Ghirardi}},
  \bibinfo{author}{\bibfnamefont{P.}~\bibnamefont{Pearle}}, \bibnamefont{and}
  \bibinfo{author}{\bibfnamefont{A.}~\bibnamefont{Rimini}},
  \bibinfo{journal}{Phys. Rev. A} \textbf{\bibinfo{volume}{42}},
  \bibinfo{pages}{78} (\bibinfo{year}{1990}).

\bibitem[{\citenamefont{Pearle}(1989)}]{Pearle1989}
\bibinfo{author}{\bibfnamefont{P.}~\bibnamefont{Pearle}},
  \bibinfo{journal}{Phys. Rev. A} \textbf{\bibinfo{volume}{39}},
  \bibinfo{pages}{2277} (\bibinfo{year}{1989}).

\bibitem[{\citenamefont{{Penrose}}(1996)}]{Penrose1996}
\bibinfo{author}{\bibfnamefont{R.}~\bibnamefont{{Penrose}}},
  \bibinfo{journal}{Gen. Relativ. Gravit.} \textbf{\bibinfo{volume}{28}},
  \bibinfo{pages}{581} (\bibinfo{year}{1996}).

\bibitem[{\citenamefont{{Di{\'o}si}}(1989)}]{Diosi1989PRA}
\bibinfo{author}{\bibfnamefont{L.}~\bibnamefont{{Di{\'o}si}}},
  \bibinfo{journal}{Phys. Rev. A} \textbf{\bibinfo{volume}{40}},
  \bibinfo{pages}{1165} (\bibinfo{year}{1989}).

\bibitem[{\citenamefont{{Hong} et~al.}(2011)\citenamefont{{Hong}, {Yang},
  {Miao}, and {Chen}}}]{Yang2011}
\bibinfo{author}{\bibfnamefont{T.}~\bibnamefont{{Hong}}},
  \bibinfo{author}{\bibfnamefont{H.}~\bibnamefont{{Yang}}},
  \bibinfo{author}{\bibfnamefont{H.}~\bibnamefont{{Miao}}}, \bibnamefont{and}
  \bibinfo{author}{\bibfnamefont{Y.}~\bibnamefont{{Chen}}},
  \bibinfo{journal}{ArXiv e-prints}  (\bibinfo{year}{2011}),
  \eprint{1110.3348}.

\bibitem[{\citenamefont{{O'Connell} et~al.}(2010)\citenamefont{{O'Connell},
  {Hofheinz}, {Ansmann}, {Bialczak}, {Lenander}, {Lucero}, {Neeley}, {Sank},
  {Wang}, {Weides} et~al.}}]{OConnell2010}
\bibinfo{author}{\bibfnamefont{A.~D.} \bibnamefont{{O'Connell}}},
  \bibinfo{author}{\bibfnamefont{M.}~\bibnamefont{{Hofheinz}}},
  \bibinfo{author}{\bibfnamefont{M.}~\bibnamefont{{Ansmann}}},
  \bibinfo{author}{\bibfnamefont{R.~C.} \bibnamefont{{Bialczak}}},
  \bibinfo{author}{\bibfnamefont{M.}~\bibnamefont{{Lenander}}},
  \bibinfo{author}{\bibfnamefont{E.}~\bibnamefont{{Lucero}}},
  \bibinfo{author}{\bibfnamefont{M.}~\bibnamefont{{Neeley}}},
  \bibinfo{author}{\bibfnamefont{D.}~\bibnamefont{{Sank}}},
  \bibinfo{author}{\bibfnamefont{H.}~\bibnamefont{{Wang}}},
  \bibinfo{author}{\bibfnamefont{M.}~\bibnamefont{{Weides}}},
  \bibnamefont{et~al.}, \bibinfo{journal}{Nature}
  \textbf{\bibinfo{volume}{464}}, \bibinfo{pages}{697} (\bibinfo{year}{2010}).

\bibitem[{\citenamefont{{Schliesser} et~al.}(2008)\citenamefont{{Schliesser},
  {Rivi{\`e}re}, {Anetsberger}, {Arcizet}, and {Kippenberg}}}]{Schliesser2007}
\bibinfo{author}{\bibfnamefont{A.}~\bibnamefont{{Schliesser}}},
  \bibinfo{author}{\bibfnamefont{R.}~\bibnamefont{{Rivi{\`e}re}}},
  \bibinfo{author}{\bibfnamefont{G.}~\bibnamefont{{Anetsberger}}},
  \bibinfo{author}{\bibfnamefont{O.}~\bibnamefont{{Arcizet}}},
  \bibnamefont{and} \bibinfo{author}{\bibfnamefont{T.~J.}
  \bibnamefont{{Kippenberg}}}, \bibinfo{journal}{Nat. Phys.}
  \textbf{\bibinfo{volume}{4}}, \bibinfo{pages}{415} (\bibinfo{year}{2008}).

\bibitem[{\citenamefont{Park and Wang}(2009)}]{Park2009}
\bibinfo{author}{\bibfnamefont{Y.-S.} \bibnamefont{Park}} \bibnamefont{and}
  \bibinfo{author}{\bibfnamefont{H.}~\bibnamefont{Wang}},
  \bibinfo{journal}{Nat. Phys.} \textbf{\bibinfo{volume}{5}},
  \bibinfo{pages}{489} (\bibinfo{year}{2009}).

\bibitem[{\citenamefont{Schliesser et~al.}(2009)\citenamefont{Schliesser,
  Arcizet, Rivi\`ere, Anetsberger, and Kippenberg}}]{Schliesser2009}
\bibinfo{author}{\bibfnamefont{A.}~\bibnamefont{Schliesser}},
  \bibinfo{author}{\bibfnamefont{O.}~\bibnamefont{Arcizet}},
  \bibinfo{author}{\bibfnamefont{R.}~\bibnamefont{Rivi\`ere}},
  \bibinfo{author}{\bibfnamefont{G.}~\bibnamefont{Anetsberger}},
  \bibnamefont{and} \bibinfo{author}{\bibfnamefont{T.~J.}
  \bibnamefont{Kippenberg}}, \bibinfo{journal}{Nat. Phys.}
  \textbf{\bibinfo{volume}{5}}, \bibinfo{pages}{509} (\bibinfo{year}{2009}).

\bibitem[{\citenamefont{Lita et~al.}(2008)\citenamefont{Lita, Miller, and
  Nam}}]{Lita2008}
\bibinfo{author}{\bibfnamefont{A.~E.} \bibnamefont{Lita}},
  \bibinfo{author}{\bibfnamefont{A.~J.} \bibnamefont{Miller}},
  \bibnamefont{and} \bibinfo{author}{\bibfnamefont{S.~W.} \bibnamefont{Nam}},
  \bibinfo{journal}{Opt. Express} \textbf{\bibinfo{volume}{16}},
  \bibinfo{pages}{3032} (\bibinfo{year}{2008}).

\bibitem[{\citenamefont{Cabrera et~al.}(1998)\citenamefont{Cabrera, Clarke,
  Colling, Miller, Nam, and Romani}}]{Cabrera1998}
\bibinfo{author}{\bibfnamefont{B.}~\bibnamefont{Cabrera}},
  \bibinfo{author}{\bibfnamefont{R.~M.} \bibnamefont{Clarke}},
  \bibinfo{author}{\bibfnamefont{P.}~\bibnamefont{Colling}},
  \bibinfo{author}{\bibfnamefont{A.~J.} \bibnamefont{Miller}},
  \bibinfo{author}{\bibfnamefont{S.}~\bibnamefont{Nam}}, \bibnamefont{and}
  \bibinfo{author}{\bibfnamefont{R.~W.} \bibnamefont{Romani}},
  \bibinfo{journal}{App. Phys. Lett.} \textbf{\bibinfo{volume}{73}},
  \bibinfo{pages}{735} (\bibinfo{year}{1998}).

\bibitem[{\citenamefont{Rempe et~al.}(1992)\citenamefont{Rempe, Thompson,
  Kimble, and Lalezari}}]{Rempe1992}
\bibinfo{author}{\bibfnamefont{G.}~\bibnamefont{Rempe}},
  \bibinfo{author}{\bibfnamefont{R.~J.} \bibnamefont{Thompson}},
  \bibinfo{author}{\bibfnamefont{H.~J.} \bibnamefont{Kimble}},
  \bibnamefont{and} \bibinfo{author}{\bibfnamefont{R.}~\bibnamefont{Lalezari}},
  \bibinfo{journal}{Opt. Lett.} \textbf{\bibinfo{volume}{17}},
  \bibinfo{pages}{363} (\bibinfo{year}{1992}).

\bibitem[{\citenamefont{Muller et~al.}(2010)\citenamefont{Muller, Flagg,
  Lawall, and Solomon}}]{Muller2010}
\bibinfo{author}{\bibfnamefont{A.}~\bibnamefont{Muller}},
  \bibinfo{author}{\bibfnamefont{E.~B.} \bibnamefont{Flagg}},
  \bibinfo{author}{\bibfnamefont{J.~R.} \bibnamefont{Lawall}},
  \bibnamefont{and} \bibinfo{author}{\bibfnamefont{G.~S.}
  \bibnamefont{Solomon}}, \bibinfo{journal}{Opt. Lett.}
  \textbf{\bibinfo{volume}{35}}, \bibinfo{pages}{2293} (\bibinfo{year}{2010}).

\bibitem[{\citenamefont{Herriott and Schulte}(1965)}]{Herriott1965}
\bibinfo{author}{\bibfnamefont{D.~R.} \bibnamefont{Herriott}} \bibnamefont{and}
  \bibinfo{author}{\bibfnamefont{H.~J.} \bibnamefont{Schulte}},
  \bibinfo{journal}{Appl. Opt.} \textbf{\bibinfo{volume}{4}},
  \bibinfo{pages}{883} (\bibinfo{year}{1965}).

\bibitem[{\citenamefont{{Jeffrey}}(2007)}]{Jeffrey2007}
\bibinfo{author}{\bibfnamefont{E.~R.} \bibnamefont{{Jeffrey}}}, Ph.D. thesis,
  \bibinfo{school}{University of Illinois at Urbana-Champaign}
  (\bibinfo{year}{2007}).

\bibitem[{\citenamefont{Vali et~al.}(1973)\citenamefont{Vali, Goldstein, and
  Fox}}]{Vali73}
\bibinfo{author}{\bibfnamefont{V.}~\bibnamefont{Vali}},
  \bibinfo{author}{\bibfnamefont{R.}~\bibnamefont{Goldstein}},
  \bibnamefont{and} \bibinfo{author}{\bibfnamefont{K.}~\bibnamefont{Fox}},
  \bibinfo{journal}{App. Phys. Lett.} \textbf{\bibinfo{volume}{22}},
  \bibinfo{pages}{391} (\bibinfo{year}{1973}).

\bibitem[{\citenamefont{Zhang et~al.}(2009)\citenamefont{Zhang, Garner, and
  Hau}}]{Zhang2009}
\bibinfo{author}{\bibfnamefont{R.}~\bibnamefont{Zhang}},
  \bibinfo{author}{\bibfnamefont{S.~R.} \bibnamefont{Garner}},
  \bibnamefont{and} \bibinfo{author}{\bibfnamefont{L.~V.} \bibnamefont{Hau}},
  \bibinfo{journal}{Phys. Rev. Lett.} \textbf{\bibinfo{volume}{103}},
  \bibinfo{pages}{233602} (\bibinfo{year}{2009}).

\bibitem[{\citenamefont{{Radnaev} et~al.}(2010)\citenamefont{{Radnaev},
  {Dudin}, {Zhao}, {Jen}, {Jenkins}, {Kuzmich}, and {Kennedy}}}]{Radnaev2010}
\bibinfo{author}{\bibfnamefont{A.~G.} \bibnamefont{{Radnaev}}},
  \bibinfo{author}{\bibfnamefont{Y.~O.} \bibnamefont{{Dudin}}},
  \bibinfo{author}{\bibfnamefont{R.}~\bibnamefont{{Zhao}}},
  \bibinfo{author}{\bibfnamefont{H.~H.} \bibnamefont{{Jen}}},
  \bibinfo{author}{\bibfnamefont{S.~D.} \bibnamefont{{Jenkins}}},
  \bibinfo{author}{\bibfnamefont{A.}~\bibnamefont{{Kuzmich}}},
  \bibnamefont{and} \bibinfo{author}{\bibfnamefont{T.~A.~B.}
  \bibnamefont{{Kennedy}}}, \bibinfo{journal}{Nat. Phys.}
  \textbf{\bibinfo{volume}{6}}, \bibinfo{pages}{894} (\bibinfo{year}{2010}).

\bibitem[{\citenamefont{Romero-Isart}(2011)}]{Romero2011}
\bibinfo{author}{\bibfnamefont{O.}~\bibnamefont{Romero-Isart}},
  \bibinfo{journal}{Phys. Rev. A} \textbf{\bibinfo{volume}{84}},
  \bibinfo{pages}{052121} (\bibinfo{year}{2011}).

\bibitem[{\citenamefont{Collett and Pearle}(2003)}]{Collett2003}
\bibinfo{author}{\bibfnamefont{B.}~\bibnamefont{Collett}} \bibnamefont{and}
  \bibinfo{author}{\bibfnamefont{P.}~\bibnamefont{Pearle}},
  \bibinfo{journal}{Found. Phys.} \textbf{\bibinfo{volume}{33}},
  \bibinfo{pages}{1495} (\bibinfo{year}{2003}).

\bibitem[{\citenamefont{{Di{\'o}si}}(2007)}]{Diosi2007}
\bibinfo{author}{\bibfnamefont{L.}~\bibnamefont{{Di{\'o}si}}},
  \bibinfo{journal}{J. Phys. A} \textbf{\bibinfo{volume}{40}},
  \bibinfo{pages}{2989} (\bibinfo{year}{2007}).

\bibitem[{\citenamefont{Maimone et~al.}(2011)\citenamefont{Maimone, Scelza,
  Naddeo, and Pelino}}]{Maimone2011}
\bibinfo{author}{\bibfnamefont{F.}~\bibnamefont{Maimone}},
  \bibinfo{author}{\bibfnamefont{G.}~\bibnamefont{Scelza}},
  \bibinfo{author}{\bibfnamefont{A.}~\bibnamefont{Naddeo}}, \bibnamefont{and}
  \bibinfo{author}{\bibfnamefont{V.}~\bibnamefont{Pelino}},
  \bibinfo{journal}{Phys. Rev. A} \textbf{\bibinfo{volume}{83}},
  \bibinfo{pages}{062124} (\bibinfo{year}{2011}).

\end{thebibliography}
\end{document}